\documentclass[aps,twocolumn,pra,superscriptaddress,amsmath,amssymb]{revtex4-1} 
\usepackage{graphicx,hyperref}
\usepackage{color}
\usepackage{dcolumn}
\usepackage{comment}
\usepackage{bm}
\usepackage{txfonts}
\usepackage[T1]{fontenc}

\begin{document}

\title{Photo-induced Dynamics of Quasicrystalline Excitonic Insulator}

\author{Ken Inayoshi}
\email{k-inayoshi@stat.phys.titech.ac.jp}
\author{Yuta Murakami}%
\author{Akihisa Koga}%
 
\affiliation{%
Department of Physics, Tokyo Institute of Technology, Meguro, Tokyo 152-8551, Japan
}%

\date{\today}
 
\begin{abstract}
  We study the photo-induced dynamics of the excitonic insulator
  in the two-band Hubbard model on the Penrose tiling
  by means of the time-dependent real-space mean-field approximation.
  We show that, with a single-cycle electric-field pulse, the bulk (spatially averaged) excitonic order parameter decreases in the BCS regime, while it increases in the BEC regime.
  To clarify the dynamics peculiar to the Penrose tiling, we examine the coordination number dependence of observables and analyze the perpendicular space.
  In the BEC regime, characteristic oscillations of the electron number at each site are induced by the pulse, which are not observed in normal crystals.
  On the other hand, the dynamics in the BCS regime is characterized by drastic change in the spatial pattern of the excitonic order parameter.
\end{abstract}

\maketitle

\section{Introduction}
By strong photo-excitations, systems can acquire novel properties~\cite{Yonemitsu2008,Aoki2013,Giannetti2016review,Basov2017review,Cavalleri2018review,Oka2019review,Sentef2021} such as nonthermal superconductivity~\cite{Fausti189,PhysRevB.89.184516,Mitrano2016,Suzuki2019} and  charge density orders~\cite{Mihailovic2014,Porer2014,Ishikawa2014,Kogar2020,Zhou2021,PhysRevB.103.054109}.
Recently, the excitonic insulating (EI) phase is attracting interests as the research target of the photo-induced nonequilibrium physics.
The EI phase is known as the macroscopic quantum condensed state of the electron-hole pairs (excitons) in the semimetals and semiconductors~\cite{Keldish1965,Rice1967EI}.
The research of the EI state has been boosted due to recent proposals of candidate materials such as  $\rm{Ta_{2}NiSe_{5}}$~\cite{PhysRevLett.103.026402,Wakisaka2012} and
$1T$-TiSe$_{2}$~\cite{PhysRevLett.99.146403,PhysRevLett.106.106404}.
Effects of strong photo-excitations on these material have been experimentally investigated,
where the enhancement~\cite{PhysRevLett.119.086401}, robustness~\cite{Baldini2021TNS} or suppression~\cite{Hellmann2012NatCom,Okazaki2018,JPSJ.89.124703,PhysRevB.103.144304} of the order have been reported depending on the excitation conditions.
These experiments stimulate further theoretical studies on nonequilibrium phenomena in the EI phase~\cite{PhysRevB.94.035121,PhysRevLett.119.247601,PhysRevB.97.115105,Tanabe2018PRB,PhysRevMaterials.3.124601,PhysRevB.101.035203,PhysRevB.102.075118,Perfetto2020PRL,Werner2020,Riku2020}.

Important questions are how the EI states respond to strong photo-excitations and how/when the EI order parameter is enhanced or suppressed.
For example, the photo-induced dynamics of the EI state has recently been examined
in the two-band Hubbard model on the normal lattice, where
a clear difference between the BCS and BEC regimes appears in the time evolution of the order parameter after the photo irradiation~\cite{PhysRevB.97.115105,PhysRevB.102.075118}.
These distinct phenomena are understood by considering the detailed dynamics of the order parameters {\it in the momentum space}.
As in this case, usually one focuses on systems on normal crystals with the translational symmetry, and the nonequilibrium phenomena are often argued in the momentum space.
On the other hand, in the solid state physics, we have a different class of materials, i.e. quasicrystals, which have ordered patterns but no translationally symmetry in the lattice~\cite{PhysRevLett.53.1951,PhysRevLett.53.2477}. 
In these systems, the analysis within the momentum space is not directly applicable.
Thus, a simple but important question arises: how is the nonequilibrium dynamics in quasicrystals  similar to or different from that in normal crystals? 

\begin{figure}
\centering
\includegraphics[width=\linewidth]{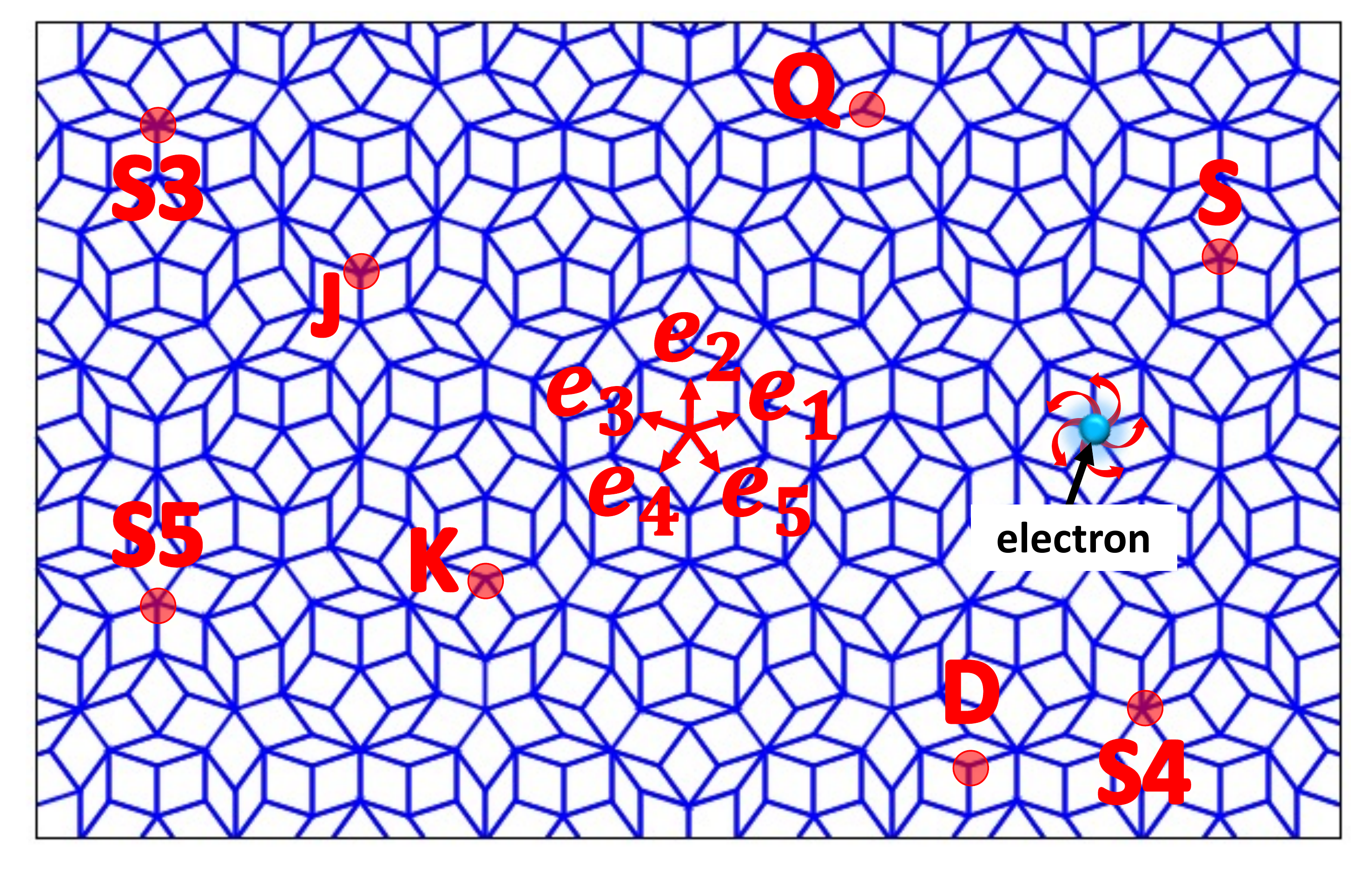}
\caption{
Vertex model on the Penrose tiling and eight types of vertices. 
$e_1, \cdots, e_5$ are projection of the translational vectors in five dimensions, ${\bf n}=(n_1,n_2,n_3,n_4,n_5)=(1,0,0,0,0),\cdots,(0,0,0,0,1)$.
Using the integers $\{n_i\}$, the lattice point $\bm{r}$ is represented as $\bm{r}=\sum_i n_i e_i$.
}
\label{Penrose}
\end{figure}

In this paper, we answer this question with respect to the photo-induced dynamics of the EI state on a quasicrystal, considering the setup similar to that for the square lattice~\cite{PhysRevB.97.115105}.
Namely, we deal with the two-band Hubbard model on the Penrose tiling~\cite{penroselattice}, which is a prototypical theoretical model of the quasicrystals, see Fig.~\ref{Penrose}.
We study this model by means of the time-dependent real-space mean-field (Hartree-Fock) approximation.
We clarify that the photo-irradiation decreases (increases) the bulk average of the EI order parameter in the BCS (BEC) regime, which phenomena are similar to that in the Hubbard model on the square lattice~\cite{PhysRevB.97.115105}.
To clarify the characteristic dynamics on the Penrose tiling, we examine the coordination number dependence of observables.
It is found that charge fluctuations are enhanced in the BEC regime, which have not been observed in the conventional periodic systems.
We also analyze the dynamics in the perpendicular space,
which allows us to discuss how the local environments affect
local physical quantities.
It is found that the spatial pattern of the EI order parameter
changes remarkably in the BCS regime.

This paper is organized as follows. 
In Sec.~\ref{sec:modelmethod}, we introduce the two-band Hubbard model
on the Penrose tiling and our numerical technique.
We briefly discuss the phase diagram in the equilibrium state.
In Sec.~\ref{sec:results},
we study the time-evolution of the system triggered by the single-cycle pulse
to clarify the dynamics peculiar to the Penrose tiling.
A summary is given in the last section.

\section{Model and Method} \label{sec:modelmethod}
We consider the two-band Hubbard model, whose Hamiltonian is expressed as
\begin{align} \label{eq:hamiltonian}
\hat{H}&=-t\sum_{\langle i,j \rangle \sigma}(\hat{c}_{i\sigma}^{\dagger}\hat{c}_{j\sigma}-\hat{f}_{i\sigma}^{\dagger}\hat{f}_{j\sigma}) 
 +\frac{D}{2}\sum_{i\sigma}(\hat{n}_{ci\sigma}-\hat{n}_{fi\sigma}) \nonumber \\
 &-\mu\sum_{i\sigma}(\hat{n}_{fi\sigma} +\hat{n}_{ci\sigma}) \nonumber \\
&+U\sum_{i}(\hat{n}_{ci\uparrow}\hat{n}_{ci\downarrow}+\hat{n}_{fi\uparrow}\hat{n}_{fi\downarrow})+V\sum_{i\sigma\sigma^{'}}\hat{n}_{ci\sigma}\hat{n}_{fi\sigma'},
\end{align}
where $\hat{f}^\dagger_{i\sigma}$ ($\hat{c}^\dagger_{i\sigma}$) is a creation operator of the electron at site $i$ with spin $\sigma\in \{ \uparrow, \downarrow \}$ in the $f$-band ($c$-band),
$\hat{n}_{ci\sigma}=\hat{c}^\dagger_{i\sigma}\hat{c}_{i\sigma}$
and $\hat{n}_{fi\sigma}=\hat{f}^\dagger_{i\sigma}\hat{f}_{i\sigma}$.
$t$ ($-t$) is the hopping integral between the nearest neighbor sites in the $f$-band ($c$-band), 
$D$ is the energy difference between two bands, and
$\mu$ is the chemical potential.
$U(>0)$ is the intraband onsite interaction
and $V(>0)$ is the interband onsite interaction.
In the following, we consider the half-filling condition i.e. $\mu=U/2+V$. 

In this paper, we treat the Penrose tiling as one of examples in quasiperiodic lattices.
It is composed of the fat and skinny rhombuses and
includes eight kinds of vertices~\cite{Bruijn1,Bruijn2},
whose coordination number (the number of bonds) takes 3 to 7,
as shown in Fig.~\ref{Penrose}.
Here, we consider the vertex model~\cite{PhysRevB.38.1621},
where electrons are located at vertices
and hop along edges of rhombuses.

To discuss photo-induced dynamics of the two-band Hubbard model on the Penrose tiling, 
we introduce the dipole transition term between two bands.
The corresponding Hamiltonian~\cite{PhysRevB.97.115105,PhysRevB.102.075118} is represented as 
\begin{align} \label{eq:dipole}
\hat{H}_{{\rm ex}}(\tau)=F_{\rm ex}(\tau)\sum_{i\sigma}(\hat{c}_{i\sigma}^{\dagger}\hat{f}_{i\sigma}+{\rm h.c.}),
\end{align}
where $F_{\rm ex}(\tau)=F_0\sin(\omega\tau)\theta(\tau)\theta(\tau_p-\tau)$ expresses the time-dependent external electric field.
 $\theta(\tau)$ is the Heaviside step function, $|F_0|$ is the magnitude of the external field, $\omega$ is the frequency, and 
$\tau_p$ is the light irradiation time.
To study the nonequilibrium dynamics,
we employ the time-dependent real-space mean-field (MF) approximation. 
This enables us to treat the large system size,
which is important to discuss electric properties
inherent in the Penrose tiling~\cite{PhysRevB.95.024509,PhysRevB.96.214402,PhysRevResearch.1.022002,Inayoshi1,PhysRevB.102.115108,PhysRevB.102.115125}.
Site-dependent MF parameters are represented
by means of the wave function $|\psi(\tau)\rangle$ as
\begin{eqnarray} 
n_{fi}(\tau)&=&\langle\psi(\tau)|\hat{f}_{i\sigma}^{\dagger}\hat{f}_{i\sigma}|\psi(\tau)\rangle, \label{OP:f}\\
n_{ci}(\tau)&=&\langle\psi(\tau)|\hat{c}_{i\sigma}^{\dagger}\hat{c}_{i\sigma}|\psi(\tau)\rangle, \label{OP:c}\\
\Delta_{i}(\tau)&=&\langle\psi(\tau)|\hat{c}_{i\sigma}^{\dagger}\hat{f}_{i\sigma}|\psi(\tau)\rangle, \label{OP:exciton}
\end{eqnarray}
where $n_{fi}(\tau)$ and $n_{ci}(\tau)$ are the electron numbers
in $f$ and $c$ bands, and
$\Delta_{i}(\tau)$ is the order parameter of the EI state at site $i$.
In the following, our discussions are restricted to the paramagnetic case, where the spin indices are omitted.
The explicit form of the MF Hamiltonian is 
\begin{align} \label{eq:TDMF}
\hat{H}_{\rm MF}^{\rm total}(\tau)&=t\sum_{\langle i,j \rangle}(\hat{f}_{i}^{\dagger}\hat{f}_{j}-\hat{c}_{i}^{\dagger}\hat{c}_{j}) +\frac{D}{2}\sum_{i}(\hat{n}_{ci}-\hat{n}_{fi}) \notag \\
&+(2V-U)\sum_{i} \left[\left\{n_{ci}(\tau)-\frac{1}{2}\right\}\hat{n}_{fi}+\left\{n_{fi}(\tau)-\frac{1}{2}\right\}\hat{n}_{ci}\right]\notag\\
&-\sum_{i}\Big[\left\{V\Delta_i(\tau)-F_{\rm ex}(\tau)\right\}\hat{f}_{i}^{\dagger}\hat{c}_{i}+{\rm h.c.}\Big].
\end{align}
Then, the time evolution of the ground state is expressed as
\begin{equation}
|\psi(\tau)\rangle=T_{\tau}{\rm exp}\left[-\frac{i}{\hbar}\int_{0}^{\tau}\hat{H}_{\rm MF}^{\rm total}(\tau') d\tau'\right]|\psi(0)\rangle,\label{wv}
\end{equation}
where $T_{\tau}$ is the time-ordering operator
and $|\psi(0)\rangle$ is the ground state of
$\hat{H}_{\rm MF}^{\rm total}(\tau=0)$.

If one examines the time evolution of the mean fields,
it is not necessary to calculate the wave function \eqref{wv}. 
Instead, we evaluate the evolution of
the single-particle density matrix defined as 
\begin{align} \label{eq:densitymatrix}
\rho_{ia,jb}(\tau)=\langle\psi(\tau)|\hat{b}_{j}^{\dagger}\hat{a}_{i}|\psi(\tau)\rangle,
\end{align}
where $\hat{a}_{i}^{\dagger}$ is a creation operator of
the electron at site $i$ and $a(=c,f)$-band.
The matrix element of the Hamiltonian ~(\ref{eq:TDMF}) is expressed in terms of
the single particle states, $|ia\rangle = \hat{a}^\dagger_i|\rm vac.\rangle$, as 
$H^{\rm MF}_{ia,jb}(\tau)=\langle ia |\hat{H}_{\rm MF}^{\rm total}(\tau)|jb\rangle$.
Then, the time evolution of $\bm{\rho}(\tau)$ is given by
\begin{align}\label{eq:Qliouville}
i\hbar\frac{\partial}{\partial \tau}\bm{\rho}(\tau)=\left[ \bm{H}^{\rm MF}(\tau), \bm{\rho}(\tau) \right].
\end{align}
Here, $\bm{H}^{\rm MF}(\tau)$ (${\bm{\rho}(\tau)}$) is the matrix with elements $H^{\rm MF}_{ia,jb}(\tau)$ ($\rho_{ia,jb}(\tau)$).
Since $\bm{H}^{\rm MF}(\tau)$ is a function of ${\bm{\rho}(\tau)}$,
we can numerically solve Eq.~(\ref{eq:Qliouville}).
Here, we use the fourth-order Runge-Kutta method with the time slice
$\Delta\tau=0.1\hbar/t$,
where the numerical error is negligible
in our simulation with $\tau<100\hbar/t$.

When no external field is applied, the electron number at site $i$ with orbital $a$ is represented as
\begin{align}
  \frac{\partial}{\partial \tau}n_{ai}(\tau)&=\frac{\partial}{\partial \tau}\rho_{ia,ia}(\tau)\nonumber \\ 
  &=-\frac{t^a}{i\hbar}
  \left\{\sum_{m}\rho_{ia,ma}(\tau)-\sum_{m}\rho_{ma,ia}(\tau)\right\},
\end{align}
where $m$ runs the nearest neighbor sites for site $i$ and $t^c=-t$, $t^f=t$.
Then, we obtain
\begin{eqnarray}
\frac{\partial}{\partial \tau}\overline{n_{a}(\tau)}=0,\label{zero} 
\end{eqnarray}
where $\overline{n_a(\tau)}=\sum_in_{ai}(\tau)/N$ and $N$ is the number of sites.
This is a natural consequence from the fact that the Hamiltonian conserves the number of electrons in each band without the electric field.
Equation~\eqref{zero} is useful to check the numerical accuracy in our simulations.

In the following, we take $t$ as the unit of the energy
and set $\hbar=1$.
Thus, the units of time and frequency are $\hbar/t=1$ and $t/\hbar=1$, respectively.
We treat the two-band Hubbard model with $U=D=4$.
The Penrose tiling is generated in terms of the deflation rule~\cite{penroselattice}.
We mainly treat the large cluster with the total sites $N=11006$ under the open boundary condition
to discuss the real-time dynamics in the quasiperiodic system.
The finite size effect will be discussed in the Appendix~\ref{sec:finite}.

\begin{figure}
\centering
\includegraphics[width=\linewidth]{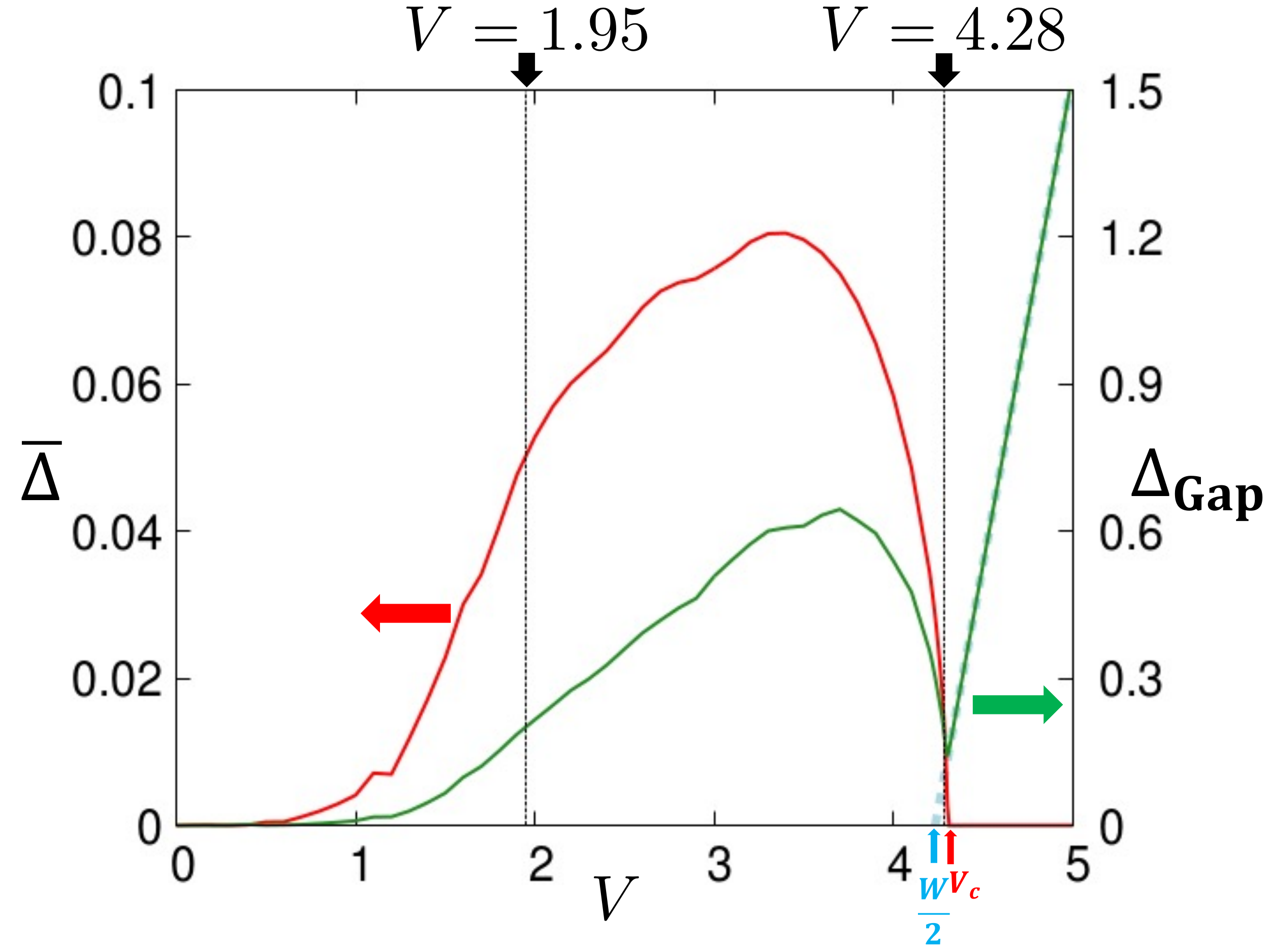}
\caption{
  $\overline{\Delta}$ and $\Delta_{\rm Gap}$ in the system with
  $U=D=4$ and $N=11006$. 
  The excitation gap in the BI state is given by $2V-W$, which is shown with the light-blue dashed line.
  Here, $W\ (\simeq8.4)$ is the bandwidth of the vertex model.
}
\label{t0delta}
\end{figure}
Before starting with discussion of the nonequilibrium dynamics,
we briefly discuss the EI state in equilibrium.
Figure~\ref{t0delta} shows the spatially-averaged EI order parameter $\overline{\Delta}$
and excitation gap $\Delta_{\rm Gap}$
as a function of the interband interaction $V$, where
$\overline{\Delta}=\sum_{i=1}^N\Delta_i(0)/N$.
Here, we take $\Delta_i(0)$ as the positive value.
The interband repulsive interaction
(electron-hole attractive interaction) widely stabilizes the EI state
against the band insulating state realized in the large $V$ region
[$V>V_c(\simeq4.3)]$.

In our study, we examine the time evolution of the EI states
in the BCS and BEC regimes to discuss the characteristic dynamics of the Penrose tiling.
We focus on the cases with the interband interactions $V=1.95$ and $4.28$
as examples of the BCS and BEC states.
In these cases, the average of the order parameter is different from each other
while the excitation gap takes the same value $\Delta_{\rm Gap}=0.2$,
as shown in Fig.~\ref{t0delta}.
\begin{figure}
\centering
\includegraphics[width=\linewidth]{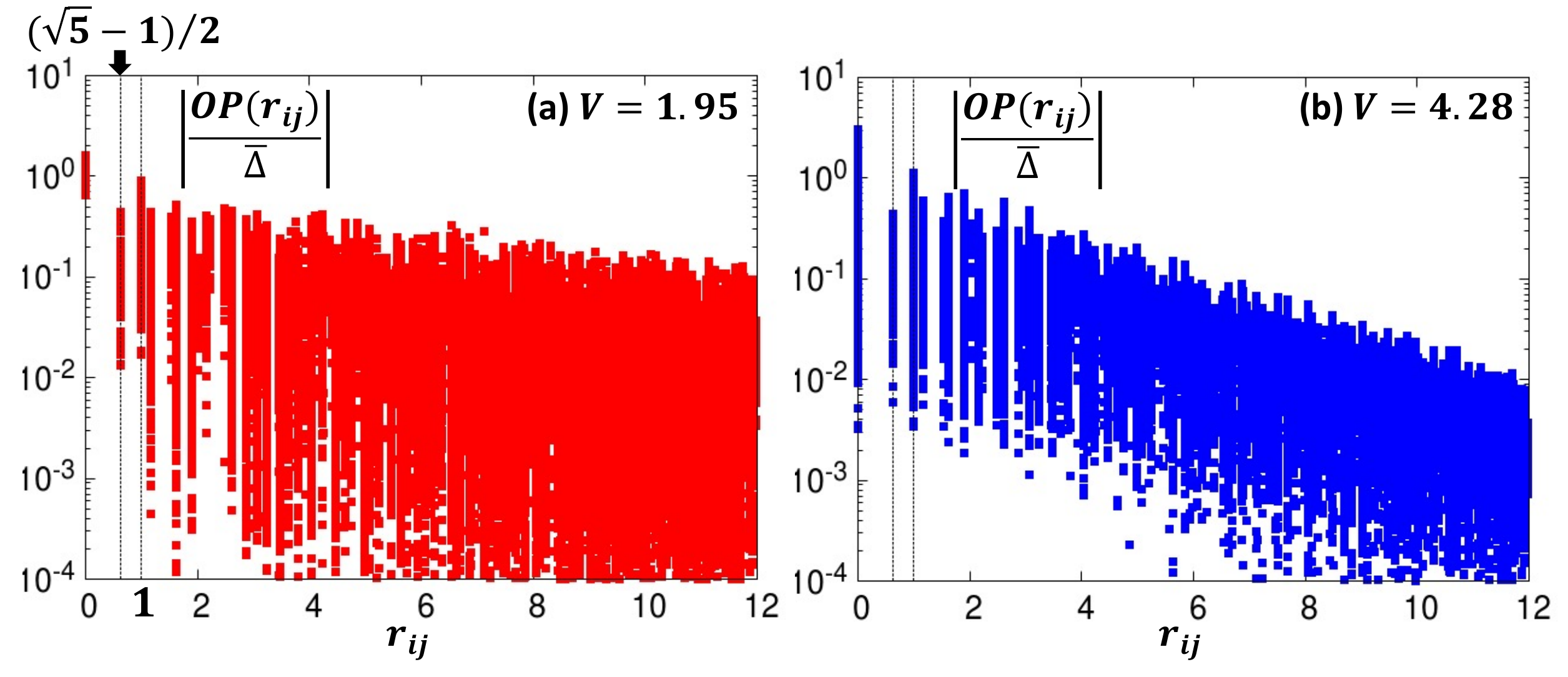}
\caption{
 Off-site electron-hole pair amplitude $OP(r_{ij})$ for the system with  $N=11006$ and $U=D=4$ 
  in (a) the BCS regime with $V=1.95$ and in (b) the BEC regime with $V=4.28$.
  Note that the shortest distance between two sites is the length of diagonal of skinny rhombus, whose length is $\frac{\sqrt{5}-1}{2}$.
}
\label{compare}
\end{figure}
A qualitative difference between the BCS and BEC regimes appears in the equilibrium.
Figure~\ref{compare} shows the off-site electron-hole pair amplitude
$OP(r_{ij})=\langle\psi_{\rm GS}|\hat{c}_i^{\dagger}\hat{f}_j|\psi_{\rm GS}\rangle$,
where $r_{ij}$ is a distance between sites $i$ and $j$.
It is found that the pair amplitude in the BEC regime decays faster
than that in the BCS regime.
This means that, in the BCS regime, the electron-hole pairs are spatially extended,
while in the BEC regime, electrons and holes are tightly coupled.
This is similar to that in normal crystals~\cite{PhysRevB.97.115105,PhysRevB.85.165135,Watanabe_2015}.
In the following, we discuss the nonequilibrium dynamics for these regimes with distinct properties.

\begin{figure*}
\centering
\includegraphics[width=0.8\linewidth]{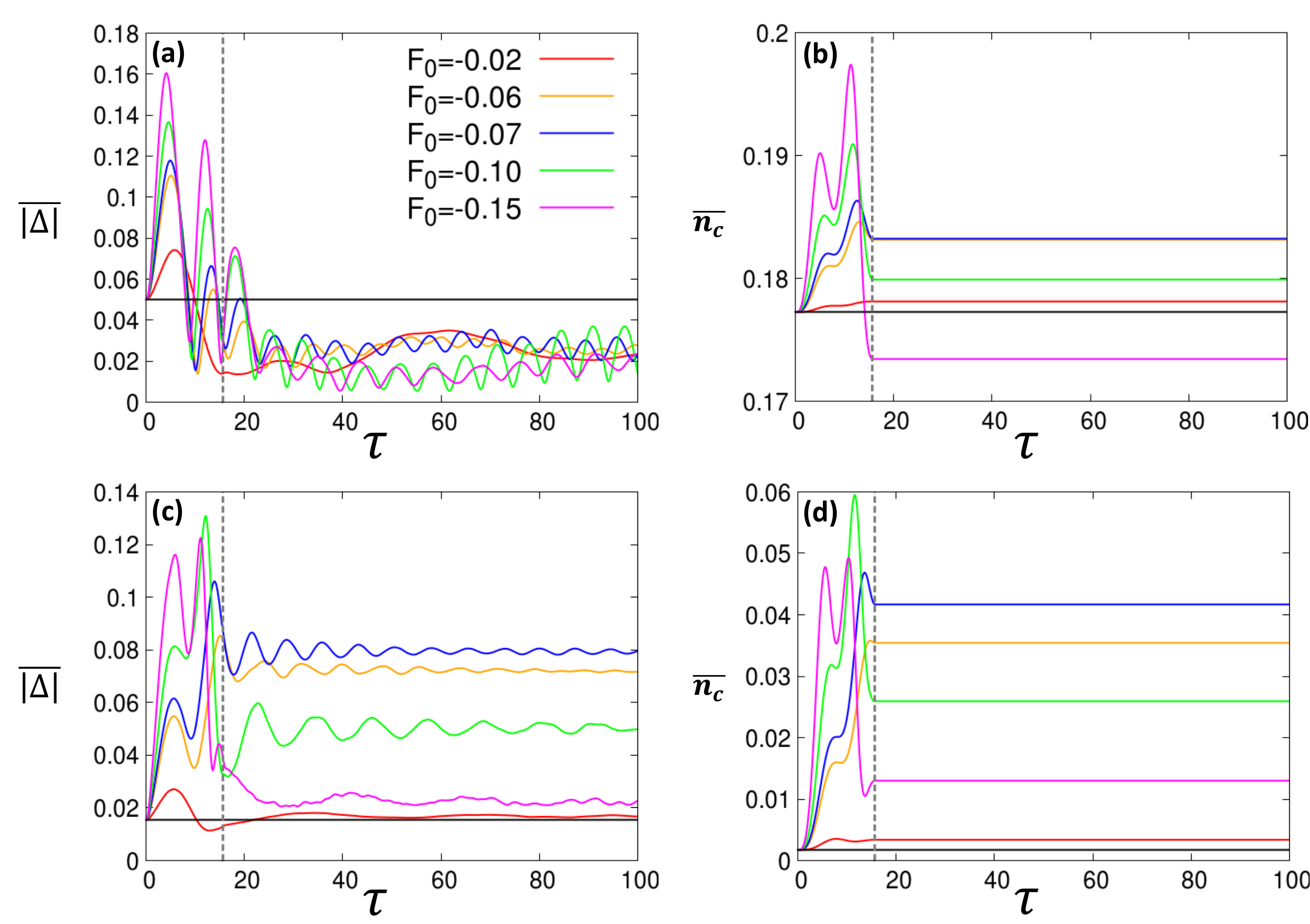}
\caption{
  Time evolutions of $\overline{|\Delta|}$ and $\overline{n_c}$
  in the system with $N=11006$ and $U=D=4$ after the single-cycle pulse is injected with several $F_0$.  
  Upper panels are the results for the BCS case with $V=1.95$,
  and lower ones are for the BEC case with $V=4.28$. 
  Horizontal black lines indicate the initial values at $\tau=0$
  and vertical dotted lines indicate $\tau_p$.
}
\label{PD}
\end{figure*}

\section{Results}\label{sec:results}
We consider the photo-induced dynamics triggered by the single-cycle pulse
with $\omega=0.4$ and $\tau_p=\frac{2\pi}{\omega}\simeq15.7$~\cite{PhysRevB.97.115105}.
Here, we set the photon energy $\omega$ twice the excitation gap $\Delta_{\rm Gap}$ so that it excites the quasiparticles with the energy beyond the gap.
Figure~\ref{PD} shows the time evolution of bulk quantities
$\overline{|\Delta(\tau)|}=\sum_{i=1}^N|\Delta_i(\tau)|/N$ and
$\overline{n_c(\tau)}=\sum_{i=1}^N n_{ci}(\tau)/N$
in the system with $U=D=4$.
These quantities are modulated by the single-cycle pulse,
and the behavior of the time evolution depends on the field strength $|F_0|$ and the interaction $V$.
We find that no oscillation appears in the electron number for each band
when $\tau>\tau_p$,
as shown in Figs.~\ref{PD}(b) and \ref{PD}(d).
This is consistent with the constraint ~(\ref{zero}), as discussed above.
By contrast, oscillatory behavior appears in the EI order parameter even when $\tau>\tau_p$,
and the frequencies of the oscillations depend on the field strength, see Figs.~\ref{PD}(a) and \ref{PD}(c).
In the BCS regime with $V=1.95$,
the EI order parameter becomes smaller than
the initial value $\overline{|\Delta(0)|}$.
On the other hand, in the BEC regime, physical quantities behave differently from those in the BCS regime
as shown in Figs.~\ref{PD}(c) and (d). 
In particular, the amplitude of the EI order parameter increases.
Similar results, i.e. the increase (decease) of the EI order parameter in the BEC (BCS) regime, have been obtained in the two-band Hubbard model on the square lattice~\cite{PhysRevB.97.115105}.
In the BCS regime, the results may seem natural since electron-hole pairs are spatially extended 
and the detail lattice structure may be less relevant for physical quantities.
In the BEC regime, when $\tau=0$, the $c$-band is almost empty and $f$-band is almost occupied.
The introduction of the single-cycle pulse rapidly increases the electron number in the $c$-band,
which leads to the formation of electron-hole pairs since the system remains coherent within the mean-field theory~\cite{ostreich1993,PhysRevLett.123.197401,PhysRevMaterials.3.124601,PhysRevB.101.035203}.
This picture is essentially the same as the explanation of the dynamics in the BEC region in the normal lattice~\cite{PhysRevB.102.075118},
which is reduced to the dynamics of a two-level system.

\begin{figure*}
\centering
\includegraphics[width=0.8\linewidth]{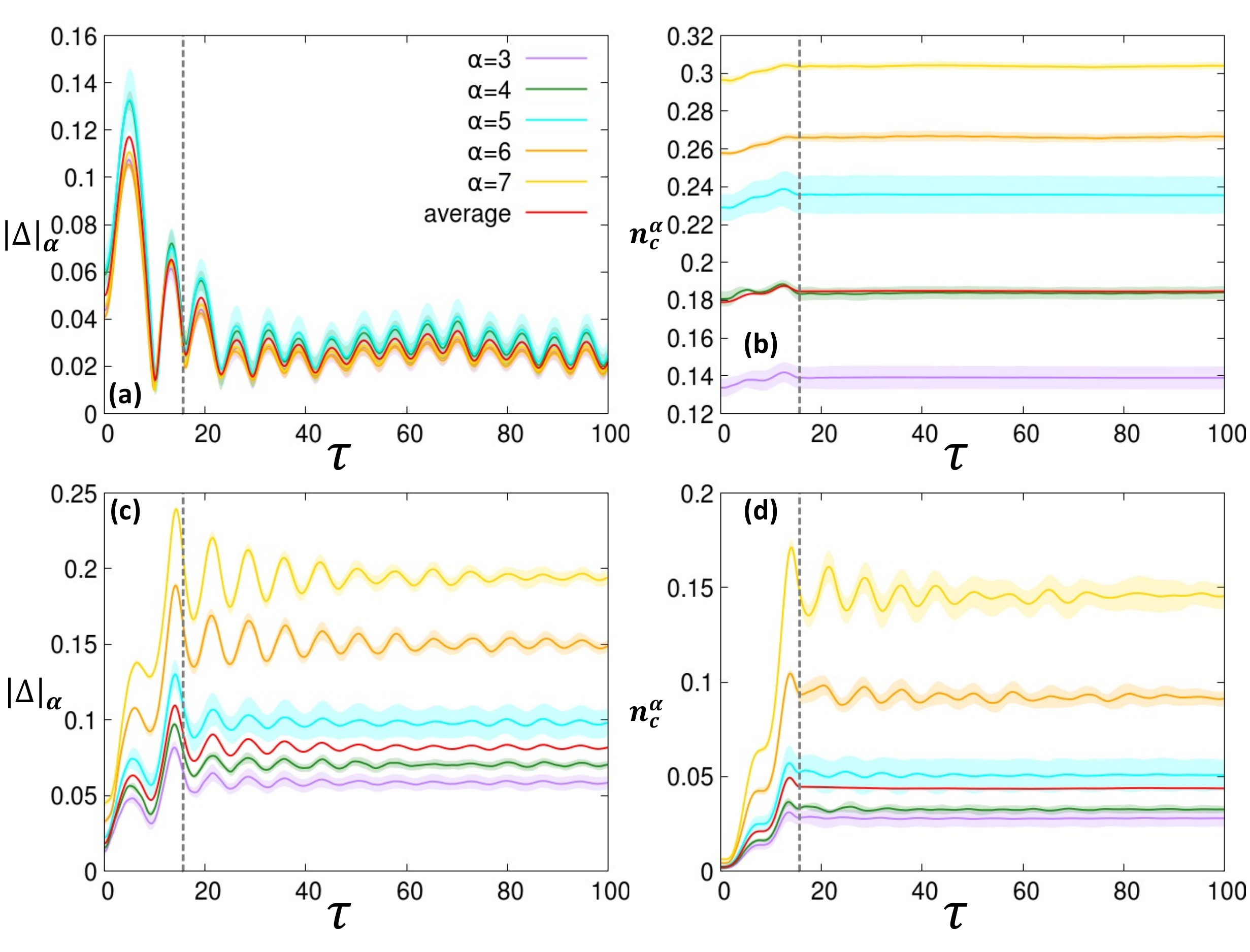}
\caption{
  Photo-induced dynamics of $|\Delta|_\alpha$ and $n_c^\alpha$ for each $\alpha$
  in the system 
  after the single-cycle pulse is injected with $F_0=-0.07$.
  Upper panels are the results for the BCS case with $V=1.95$, and
  lower ones are for the BEC case with $V=4.28$.
  Red lines represent $\overline{|\Delta|}'$ and $\overline{n_c}'$,
  and vertical dotted lines represent $\tau_p$.
  Note that $\overline{n_c}'$ oscillates slightly although $\overline{n_c}$ is conserved when $\tau>\tau_p$,
  see Fig.~\ref{7Gvs8G} for an enlarged view of $\overline{n_c}'$.
}
\label{8Geach}
\end{figure*}

So far, we showed that the qualitative behavior of the spatially-averaged quantities is similar to that in normal crystals.
We now focus on the spatial dependence of physical quantities and reveal the effects of the quasiperiodic structure
in the nonequilibrium dynamics.
One of the important features of the Penrose tiling is that
the coordination number at site $i$ $Z_i$ takes 3 to 7, in contrast to the square lattice.
In the following, to avoid the boundary effects in the system, we consider the {\it bulk} region.
The definition of it is explicitly shown in the Appendix~\ref{sec:noboundary}.
The bulk region includes $N'=7936$ sites when one treats the system with $N=11006$.
To see the coordination dependence of physical quantities, we introduce the coordination-dependent averages as 
\begin{eqnarray}
  |\Delta(\tau)|_\alpha&=&\frac{1}{N'_\alpha}\sum_{i\ \text{with}\ Z_i=\alpha}|\Delta_i(\tau)|,\\
  n_c^\alpha(\tau)&=&\frac{1}{N'_\alpha}\sum_{i\ \text{with}\ Z_i=\alpha}n_{ci}(\tau),
\end{eqnarray}
where $N'_\alpha$ is the number of the lattice sites with $Z_i=\alpha\ (3,\cdots,7)$ in the bulk region.
Figure~\ref{8Geach} shows the results for the system with $F_0=-0.07$,
where the standard deviations of the quantities are drawn as the shaded areas.
We also plot averaged values $\overline{|\Delta(\tau)|}'=\sum_{\alpha}N'_\alpha|\Delta(\tau)|_\alpha/N'$ and $\overline{n_c(\tau)}'=\sum_{\alpha}N'_\alpha n_c^\alpha(\tau)/N'$.
We find that $|\Delta_i(\tau)|$ and $n_{ci}(\tau)$ are well
classified by the coordination number 
although $|\Delta(\tau)|_\alpha$ behaves qualitatively in the similar way to its spatial average $\overline{|\Delta|}$,
as shown in Fig.~\ref{PD}.
Namely, in the BCS regime,
the EI order parameter decreases by the single-cycle pulse
and it increases in the other.
We also note that the frequency for oscillatory behavior in the EI order parameter
does not depend on the coordination number.
This may be trivial in the BCS regime since
electron-hole pairs are spatially extended
and  physical quantities do not strongly depend on the vertices.
On the other hand, in the BEC regime, the electron-hole pairs are tightly coupled and thus the excitonic properties should be mainly determined by local structures (vertices).
Figures ~\ref{8Geach}(c) and \ref{8Geach}(d) show that the nonequilibrium behavior of physical quantities is well categorized by the coordination number. 
Such a vertex dependence of the physical quantities is one of the features in the quasicrystalline systems.
Thus, it may be nontrivial that there is only a small difference in the frequency of the oscillations, see Fig. ~\ref{8Geach}(c).

\begin{figure}
\centering
\includegraphics[width=\linewidth]{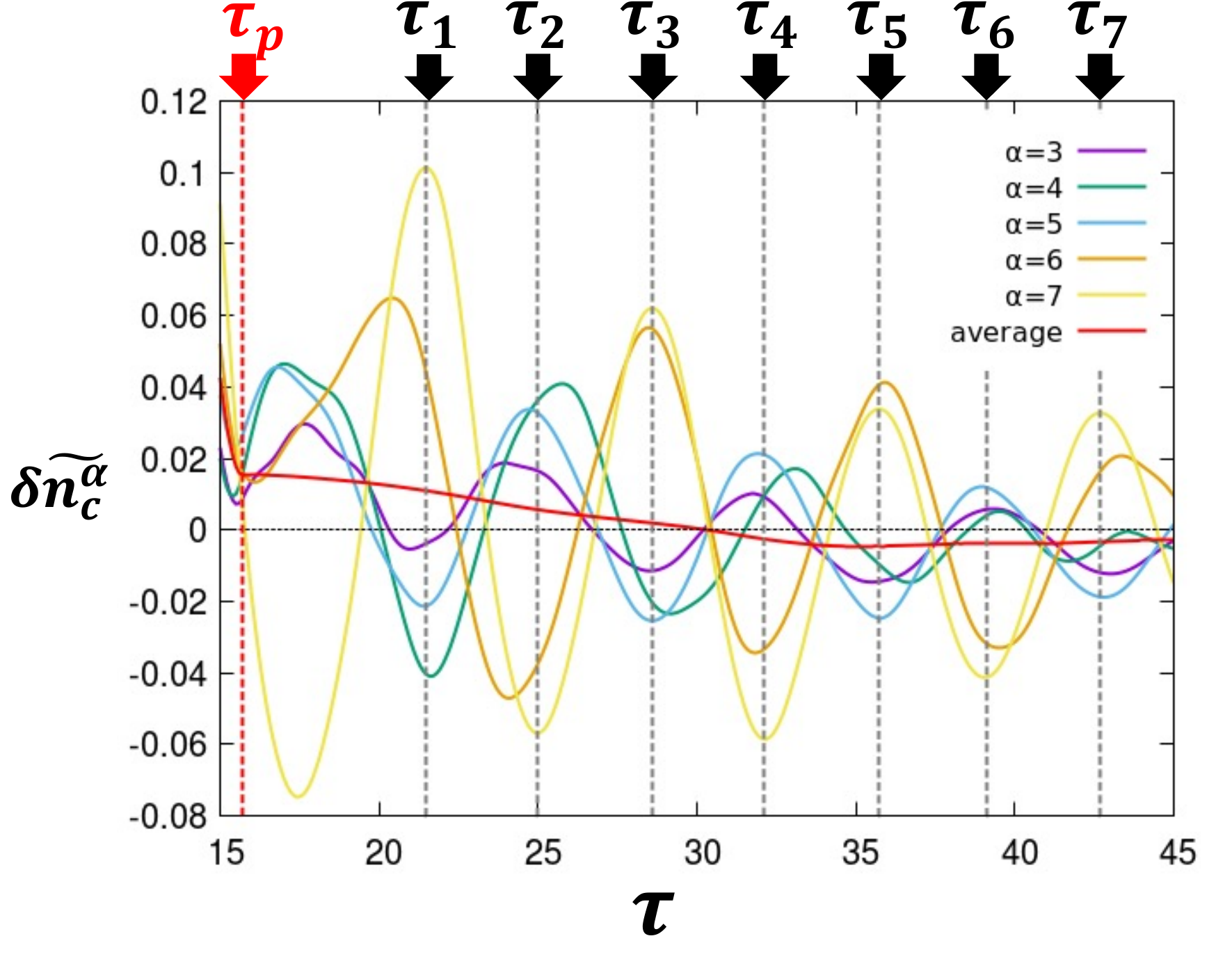}
\caption{
  Photo-induced dynamics of $\delta\widetilde{n_c^\alpha}(\tau)$ for each $\alpha$ in the BEC case with $V=4.28$
  after the single-cycle pulse is injected with $F_0=-0.07$.
  Red line represents $\delta\widetilde{n_c}'(\tau)$.
}
\label{ncfluc}
\end{figure}

We also note that  oscillatory behavior appears in the electron number $n_c^\alpha$ at $\tau>\tau_p$
although their total number $\overline{n_c}$ is always constant, see Fig.~\ref{8Geach}(d).
Since such peculiar charge fluctuations are trivially absent in the normal crystals and are not visible in the BCS regime with $V=1.95$,
they are  characteristic dynamics of the BEC regime in the Penrose tiling.
To look in detail the oscillatory behavior in $n_c^\alpha$ for the sites with $Z_i=\alpha$,
we introduce the deviation from the time average as,
\begin{align}
\delta\widetilde{n_c^\alpha}(\tau)=\frac{n_{c}^{\alpha}(\tau)}{\widetilde{n_c}(\alpha)}-1,
\end{align}
where $\widetilde{n_c}(\alpha)(=\frac{1}{\tau_7-\tau_1}\int_{\tau_1}^{\tau_7}n_c^\alpha(\tau) \ d\tau)$ is an average in the interval $(\tau_1, \tau_7)$,
and $\tau_i$ is the maximum or minimum in the curve of $n_c^7(\tau)$ with $\tau>\tau_p$, see Fig.~\ref{ncfluc}.
We also plot $\delta\widetilde{n_c}'(\tau)=\overline{n_c}'(\tau)/\widetilde{n_c}'-1$ where $\widetilde{n_c}'(=\frac{1}{\tau_7-\tau_1}\int_{\tau_1}^{\tau_7}\overline{n_c}'(\tau) \ d\tau)$.
The results are shown in Fig.~\ref{ncfluc}.
It is found that the charge oscillation induced by the single-cycle pulse decays with increasing $\tau$.
The small change of $\delta\widetilde{n_c}'$ ($\overline{n_c}'$) is caused by the finite size effect, see the Appendix~\ref{sec:finite}.
We note that the quantities $n_c^\alpha$ can be classified into two groups
$\{n_c^6, n_c^7\}$ and $\{n_c^3, n_c^4,n_c^5\}$,
where the relative phase of their oscillations is almost $\pi$.
This difference is consistent with the fact that
the total number of electrons in $c$-band never changes
when $\tau>\tau_p$.
We note that the total number of electrons at each site ($n_{ci}+n_{fi}$) remains unity during the time evolution. 
Therefore, the charge oscillation is distinct from a charge density wave induced by the pulse.

\begin{figure*}
\centering
\includegraphics[width=0.9\linewidth]{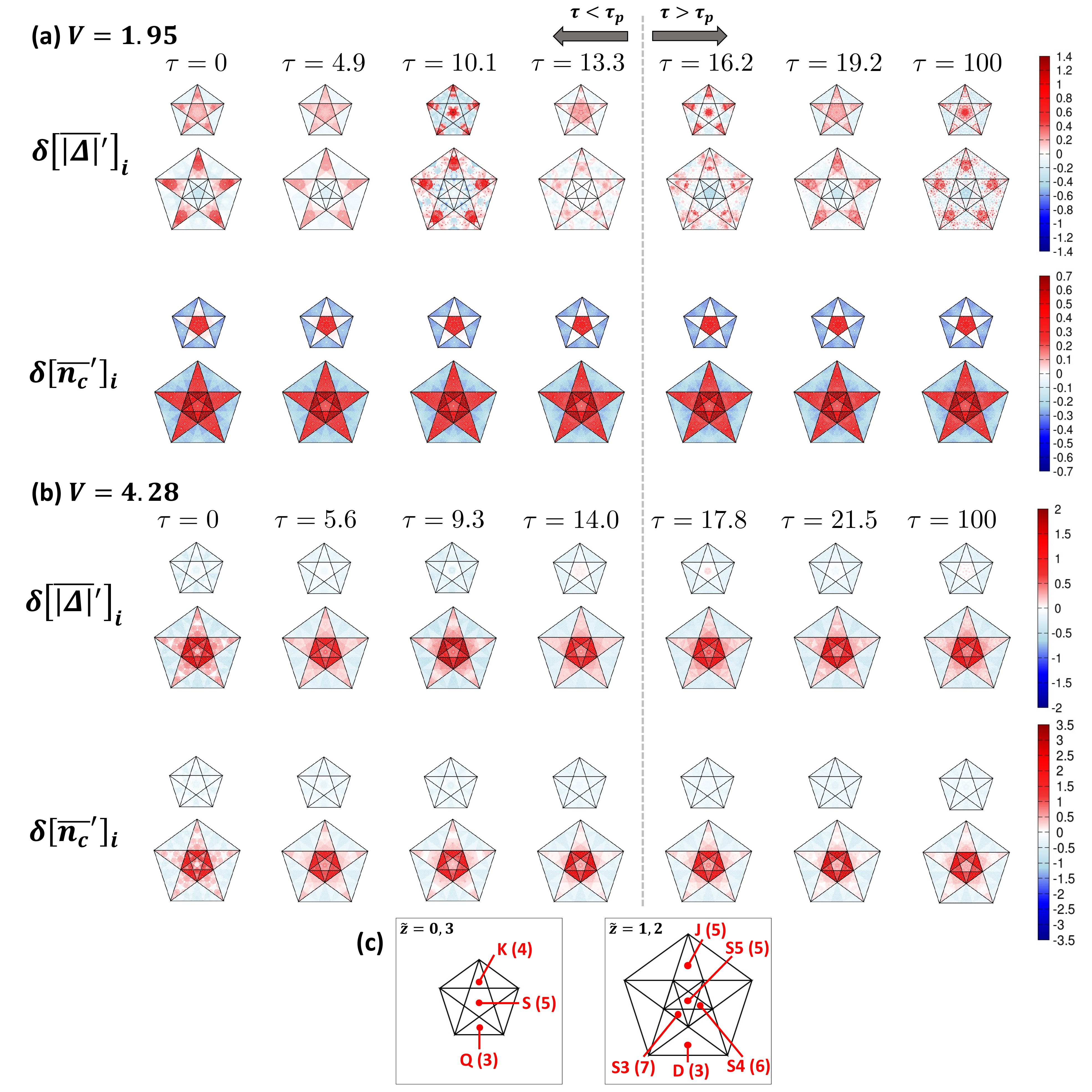}
\caption{
  Time-evolution of $\delta\left[\overline{|\Delta|}'\right]_i$ and $\delta\left[\overline{n_c}'\right]_i$ in the perpendicular space
  for the (a) BCS ($V=1.95)$ and (b) BEC ($V=4.28$) cases with $F_0=-0.07$. 
  (c) Each domain is the region for the eight kinds of vertices
  shown in Fig.~\ref{Penrose}. 
  The integer in parenthesis indicates the coordination number for each vertex.
}
\label{perD}
\end{figure*}

\begin{figure*}
\centering
\includegraphics[width=0.9\linewidth]{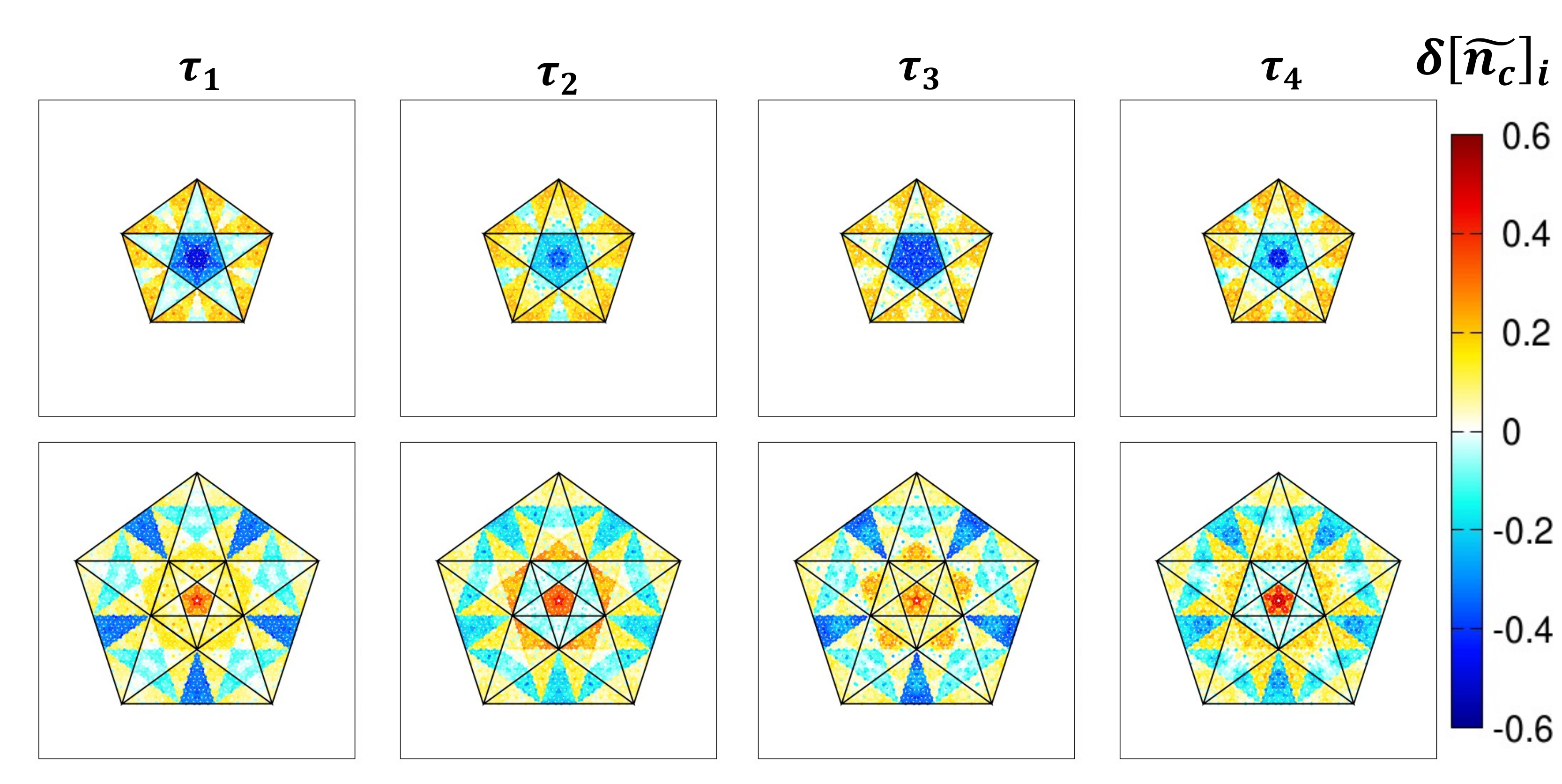}
\caption{
  Time-evolution of local charge fluctuations $\delta\left[\widetilde{n_c}\right]_{i}$ in the perpendicular space
  for the BEC case ($V=4.28$) with $F_0=-0.07$.
}
\label{ncfluc2}
\end{figure*}

Now we discuss the site dependence of physical quantities from a bit different point of view using the perpendicular space~\cite{perpendicularspace}.
This space is useful since it allows us to systematically discuss how the local lattice structures affect the physical quantities.
On the Penrose tiling, each site is represented 
by the five dimensional vector ${\bf n} = (n_1,n_2,n_3,n_4,n_5)$
with integers $n_{\mu}$ as shown in Fig.~\ref{Penrose}. 
Its coordinate $\bm{r}$ is constructed
by the projection onto the two dimensions as,
\begin{equation}
\label{eq:r}
\bm{r}=(x,y)=({\bf n}\cdot{\bf e}^x,{\bf n}\cdot{\bf e}^y),
\end{equation}
where $e^{x}_{\mu} = \cos(\phi\mu+\theta_0)$,
$e^{y}_{\mu} = \sin(\phi\mu+\theta_0)$, and $\phi = 2\pi/5$.
The initial phase $\theta_0$ is arbitrary and
we set $\theta_0=-\frac{3\pi}{10}$ as an example.
The projection onto the three-dimensional perpendicular space is given by
\begin{eqnarray}
\label{eq:rperp}
\tilde{\bf r}&=&(\tilde{x},\tilde{y})=({\bf n}\cdot\tilde{\bf e}^x, {\bf n}\cdot\tilde{\bf e}^y),\\
\tilde{z}&=&{\bf n}\cdot\tilde{\bf e}^z,
\end{eqnarray}
where $\tilde{e}^{x}_{\mu} = \cos(2\phi\mu+\theta_0)$,
$\tilde{e}^{y}_{\mu} = \sin(2\phi\mu+\theta_0)$, and
$\tilde{e}^{z}_{\mu} = 1$.
It is known that $\tilde{z}$ takes only four consecutive integers.
In each $\tilde{z}$-plane, the $\tilde{\bf r}$-points densely cover
a region of pentagon shape.
The pentagon in $3-\tilde{z}$ plane has the same size
as the pentagon in $\tilde{z}$ plane.
Eight kinds of vertices,
which are quasiperiodically arranged in the real space as Fig.~\ref{Penrose},
are mapped to distinct domains, as shown in Fig.~\ref{perD}(c).
Therefore, this perpendicular-space analysis allows us to
discuss how site-dependent physical quantities are characterized
by the local lattice structures, which include more information than the coordination number.

Here, we calculate the deviation of the quantities,
\begin{align}
\delta\left[\overline{|\Delta|}'\right]_i(\tau)&=\frac{|\Delta_i(\tau)|}{\overline{|\Delta(\tau)|}'}-1, \\
\delta\left[\overline{n_c}'\right]_i(\tau)&=\frac{n_{ci}(\tau)}{\overline{n_c(\tau)}'}-1,
\end{align}
and we show the results in Fig.~\ref{perD}.
Now, we plot $\delta\left[\overline{|\Delta|}'\right]_i$ and $\delta\left[\overline{n_c}'\right]_i$ on $\tilde{z}$ and $3-\tilde{z}$ planes on the same plane because the profiles for $\tilde{z}$ and $3-\tilde{z}$ planes are identical in the thermodynamic limit ($N\rightarrow\infty$).
When the system belongs to the BCS regime with $V=1.95$,
the average of $n_c^\alpha$ is little changed by the time evolution,
as shown in Fig.~\ref{8Geach}(b).
This is also found in the perpendicular space,
where $\delta\left[\overline{n_c}'\right]_i$ is almost constant in each domain for the corresponding vertex,
as shown in Fig.~\ref{perD}(a).
On the other hand, different behavior appears
in the distribution of $|\Delta_i|$.
For example, we focus on the $D$ and $J$ vertices.
In the initial state with $\tau=0$,
the magnitude of their order parameters is smaller (larger) than the total average on the  the $D$ ($J$) vertices,
which is clearly shown in the corresponding domains.
After the single-cycle pulse is injected, 
we find red and blue regions in the $D$ and $J$ domains, which implies that 
the oscillatory behavior in the order parameter is not specified by
the kinds of vertices,
in contrast to the charge distribution.
This distinct behavior is characteristic of the BCS regime.
By contrast, in the BEC case with $V=4.28$,
the distributions of $\delta\left[\overline{|\Delta|}'\right]_i$ and
$\delta\left[\overline{n_c}'\right]_i$ have a similar structure in the perpendicular space.
These indicate that, in the BEC case,
the system is mainly described only by the local lattice structures.
Nevertheless, in the case, charge fluctuations
are induced by the injection of the single-cycle pulse, as discussed above.

To see the spatial pattern of the charge fluctuations, we show $\delta\left[\widetilde{n_c}\right]_{i}(\tau)=n_{ci}(\tau)/\widetilde{n_c}(Z_i)-1$
for the BEC state ($V=4.28$) with $F_0=-0.07$ in Fig.~\ref{ncfluc2}.
In the domains for $S4$ and $S3$ vertices where $\alpha=6,7$ in the perpendicular space,
we find that the quantities clearly oscillate together with sign changes.
By contrast, in the other domains for $\alpha=3, 4$, and $5$, 
we could not see clear oscillatory behavior with sign changes.
In addition, we find that the domain can be further classified into some subdomains.
For examples, the $D$ domain is split into seven subdomains,
as shown in Fig.~\ref{ncfluc2}.
These two points are consistent with the fact that the width of oscillations is smaller than
its standard deviation, as shown in Fig.~\ref{8Geach}(d).
The existence of subdomain structures implies that the local charge fluctuations are affected by
not only the coordination number but also the environment of the connecting sites.
In fact, such a subdomain structure in $\delta\left[\widetilde{n_c}\right]_{i}$ is not changed
during the time evolution.

We wish to note that even when the initial state is in the band insulating state with $\Delta_i=0$ and $V>V_c$,
substantial size of the excitonic order parameter appears due to the single-cycle pulse and 
oscillatory behavior similar to the BEC regime emerges (not shown).
This implies the existence of photo-induced transient EI order~\cite{PhysRevLett.123.197401} in the quasicrystal,
and our results may be relevant for dynamics of photo-excited semiconductors.
Although neither of an excitonic insulator or a semiconductor on a quasicrystal has been found up to now,
the semiconducting approximant Al-Si-Ru has recently been synthesized~\cite{PhysRevMaterials.3.061601}.
We believe that semiconducting quasicrystals will be synthesized in near future,
and interesting excitonic properties discussed here should be observed.

\section{SUMMARY AND OUTLOOK}
In this study, we have examined the photo-induced dynamics of
the EI phase in the two-band Hubbard model on the Penrose tiling.
It is found that after the single-cycle pulse is injected
the magnitude of the EI order parameters decreases in the BCS regime and it increases in the BEC regime, which is similar to that in the conventional periodic systems.
Furthermore we have discussed nonequilibrium phenomena peculiar to the Penrose tiling.
Examining the coordination number dependence in the physical quantities,
we have found oscillatory behavior of the $c$-band electron number.
Since the charge oscillation is not prominent in the BCS regime, the induced charge fluctuations are inherent in the BEC regime.
We further clarified the difference of the dynamics between the BCS and BEC regimes in terms of the perpendicular space analysis.
In the BEC regime, the patterns of the EI order parameter and the number of $c$-band electron are similar, which holds even after the photo excitation.
On the other hand, in the BCS regime, the pattern of the EI order parameter is distinct from that of the $c$-band electron number.
In particular, the pattern of the order parameters changes remarkably with the photo excitation.

We believe that the nonequilibrium dynamics in quasicrystal systems hosts potentially interesting questions and our work should be a milestone for researches in this direction.
One of interesting topics in the field is the role of the confined states, which are macroscopically degenerate states peculiar to quasicrystals.
In our previous work~\cite{Inayoshi1}, it has been found that the EI order parameter shows intriguing spacial distribution reflecting the confined states.
It is interesting and important to clarify that this unique distribution can be photo-induced or changed in response to the photo irradiation.
These are now under consideration.

\appendix
\section{Bulk Region in Our Model}\label{sec:noboundary}
\begin{figure}
\centering
\includegraphics[scale=0.3]{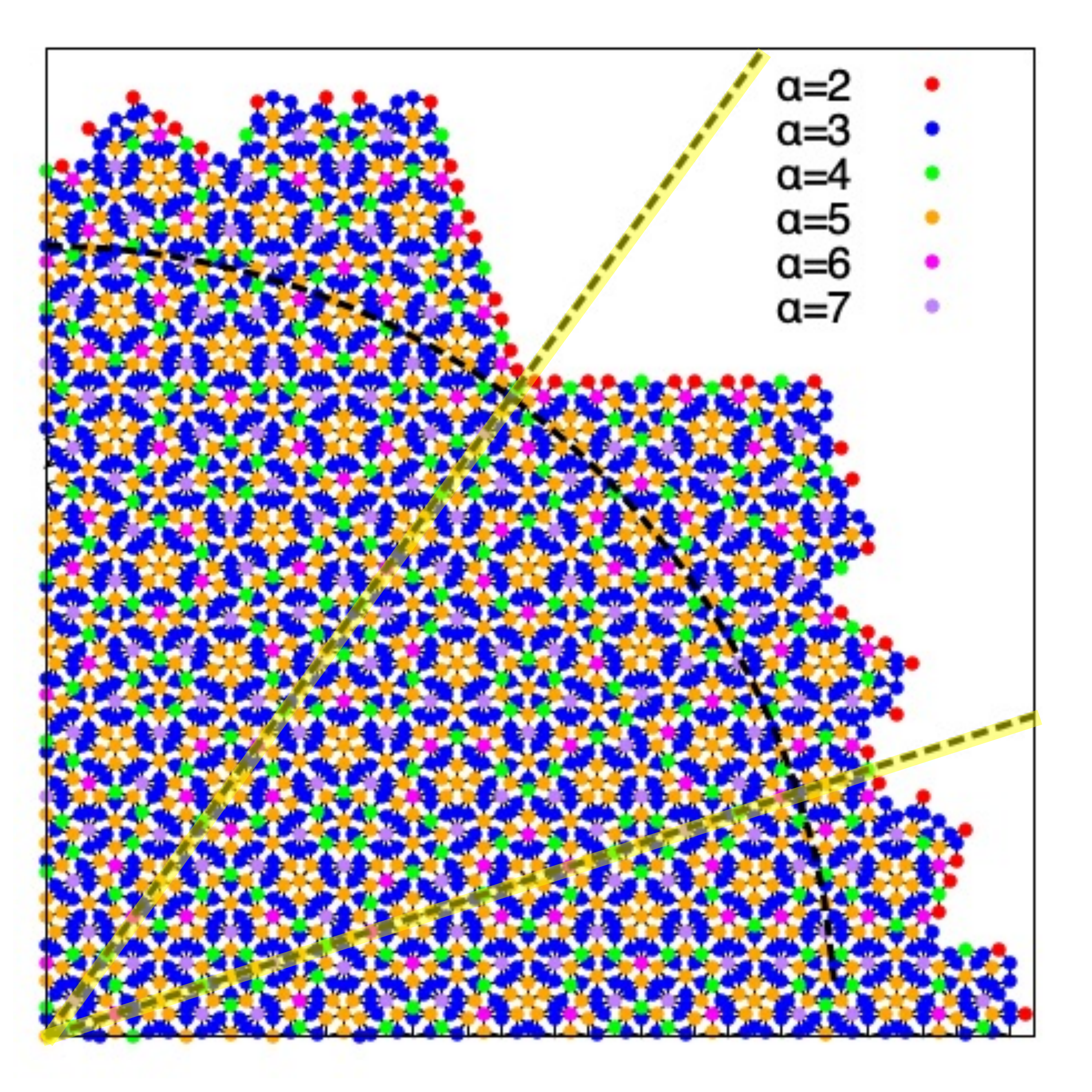}
\caption{The area within the dashed arc line indicates the bulk region.
The area between two yellow lines in the figure is one of the ten equivalent regions defined by the $C_{5v}$ symmetry 
of the Penrose tiling.
}
\label{noboundaryregion}
\end{figure}

In order to eliminates the effects of the edges, we define the {\it bulk} region as an area within a reasonable distance  from the center of the tiling.
Specifically, we take a circular area as shown in Fig.~\ref{noboundaryregion}.
The system has the $C_{5v}$ symmetry and it can be separated into ten equivalent regions,
one of which is the area between two yellow lines in Fig.~\ref{noboundaryregion}.
The area inside the black dashed arc is taken as the bulk region in the $N=11006$ system we used.
When we denote the number of sites with the coordination number $\alpha$ in the whole system and in the bulk region as $N_{\alpha}$ and $N_{\alpha}'$, respectively, we have $(N_2,N_3,N_4,N_5,N_6,N_7)=(180,5795,995,3066,405,565)$ and $(N_2',N_3',N_4',N_5',N_6',N_7')=(0,4195,725,2296,275,445)$.

\section{The Effects of the System Size and Edges}\label{sec:finite}
\begin{figure}[h]
\centering
\includegraphics[width=\linewidth]{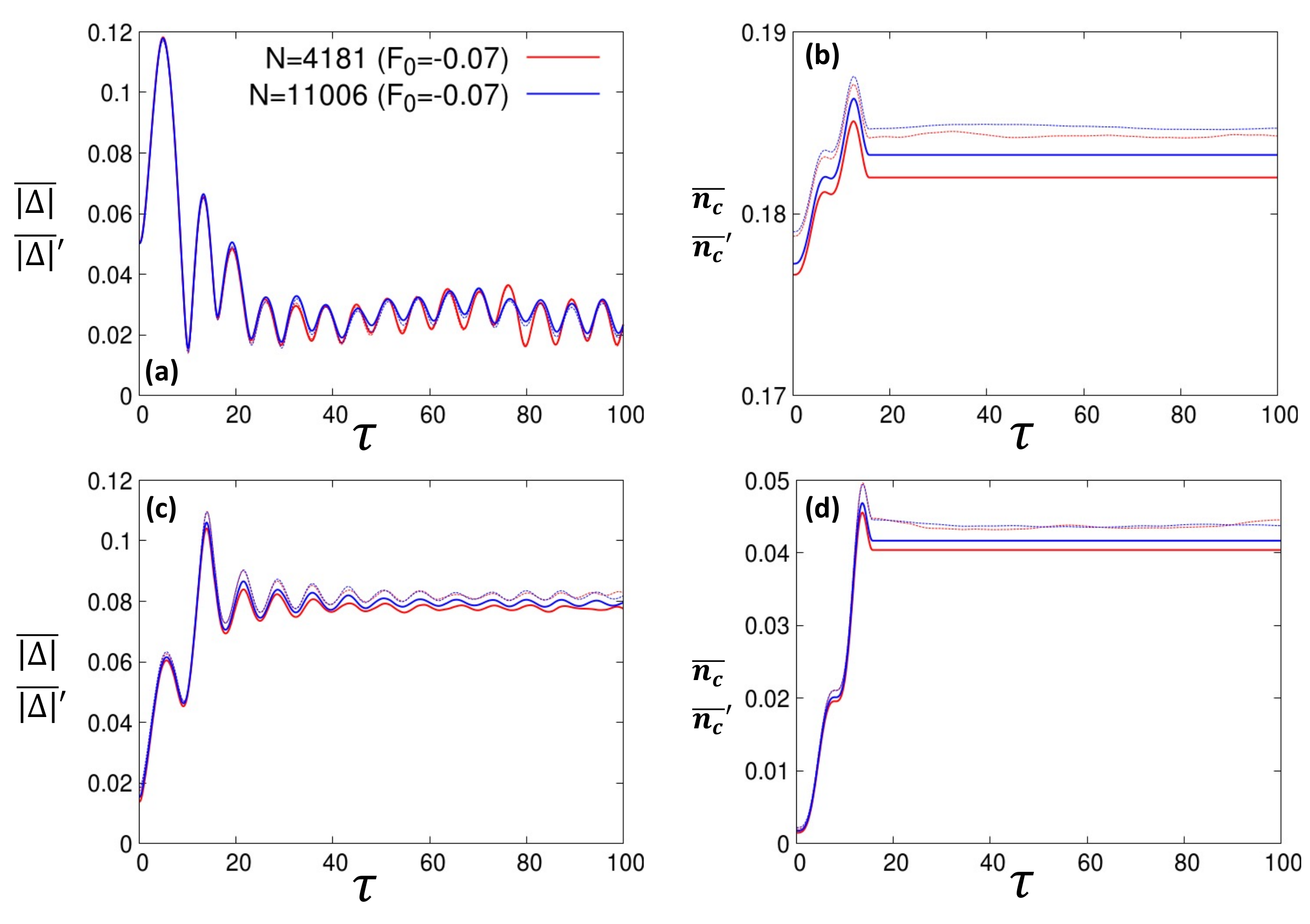}
\caption{Comparison of photo-induced dynamics between the system with $N=11006$ and the system with $N=4181$ under the condition $(U,D,\omega,F_0)=(4,4,0.4,-0.07)$. 
(a)(b) Time evolution of $\overline{|\Delta|}$ and $\overline{n_c}$ for $V=1.95$.
(c)(d) Time evolution of $\overline{|\Delta|}$ and $\overline{n_c}$ for $V=4.28$. 
Dashed lines represent $\overline{|\Delta|}'$ and $\overline{n_c}'$ for each system.}
\label{7Gvs8G}
\end{figure}

To discuss the effects of the system size and edges, we look at the dynamics of $\overline{|\Delta|}$, $\overline{|\Delta|}'$, $\overline{n_c}$, and $\overline{n_c}'$ 
under the conditions, $(U,D,V,\omega,F_0)=(4,4,1.95,0.4,-0.07)$ and $(U,D,V,\omega,F_0)=(4,4, 4.28,0.4,-0.07)$, in the system with $N=11006\ (N'=7936)$ and  the system with $N=4181\ (N'=2921)$, see Fig.~\ref{7Gvs8G}.
It is found that the qualitative behavior of $\overline{|\Delta|}$, $\overline{|\Delta|}'$, $\overline{n_c}$, and $\overline{n_c}'$ is similar in the systems with $N=11006$ and $N=4181$.
However, strictly speaking, the detailed values of $\overline{|\Delta|}$, $\overline{|\Delta|}'$, $\overline{n_c}$, and $\overline{n_c}'$ are different.
If we want to evaluate the accurate values in the thermodynamic limit, we need to calculate the time evolution for larger systems, 
which is too expensive for the current computational resources. 
Therefore, in this paper, we focus on the qualitative aspects. 

\begin{acknowledgements}
We thank K. Yonemitsu for fruitful discussions.
Parts of the numerical calculations are performed in the supercomputing systems in ISSP,
the University of Tokyo.
K.I. acknowledges the financial supports from 
Advanced Research Center for Quantum Physics and Nanoscience, 
and Advanced Human Resource Development Fellowship for Doctoral Students, Tokyo Institute of Technology.
This work was supported by a Grant-in-Aid for Scientific Research from JSPS,
KAKENHI Grant Nos. JP19K23425, JP20K14412, JP20H05265 (Y.M.),
JP21H01025, JP19H05821, JP18K04678, JP17K05536 (A.K.),
JST CREST Grant No. JPMJCR1901 (Y.M.).
\end{acknowledgements}

\bibliography{refs}

\end{document}